\documentclass[twocolumn,showpacs,preprintnumbers,amsmath,amssymb]{revtex4}

\usepackage{graphicx}
\usepackage{dcolumn}
\usepackage{bm}
\usepackage{paralist}
\usepackage{amsmath}
\usepackage{epsfig}

\newcommand{\vex}[1]{\mathbf{#1}}
\newcommand{\ten}[1]{\mathbf{#1}}
\newcommand{\tentau}{\mathbf{T}}
\newcommand{\vnabla}{\boldsymbol{\nabla}}

\newcommand{\pdert}[1]{\frac{\partial #1}{\partial t}}

\newcommand{\eqnref}[1]{Eq.~(\ref{#1})}
\newcommand{\figref}[1]{Fig.~\ref{#1}}

\newcommand{\etal}[1]{\textit{et al.}}
\def\square{${\vcenter{\hrule height .8pt
        \hbox{\vrule width .8pt height 5pt \kern 5pt
        \vrule width .8pt}
        \hrule height .8pt}}$}

\newcount\ndots
\def\drawline#1#2{\raise 2.5pt\vbox{\hrule width #1pt height #2pt}}
\def\spacce#1{\hskip #1pt}
\def\solid{\drawline{24}{.5}\nobreak\ }
\def\boldsolid{\drawline{24}{1.}\nobreak\ }
\def\bdash{\hbox{\drawline{4}{.5}\spacce{2}}}
\def\dashed{\bdash\bdash\bdash\bdash\nobreak\ }
\def\boldbdash{\hbox{\drawline{4}{1.}\spacce{2}}}
\def\bolddashed{\boldbdash \boldbdash \boldbdash \boldbdash\nobreak\ }
\def\bdot{\hbox{\drawline{1}{.5}\spacce{2}}}
\def\dotted{\hbox{\leaders\bdot\hskip 24pt}\nobreak\ }

\def\trian{\raise 1.25pt\hbox{$\scriptscriptstyle\triangle$}\nobreak\ }
\def\circle{$\circ$\nobreak\ }

\def\square{${\vcenter{\hrule height .4pt
        \hbox{\vrule width .4pt height 3pt \kern 3pt
        \vrule width .4pt}
        \hrule height .4pt}}$\nobreak\ }
\def\plus{\raise 1.25pt \hbox{$\scriptscriptstyle +$}\nobreak\ }
\def\x{\raise 1.25pt \hbox{$\scriptscriptstyle \times$}\nobreak\ }

\def\solidtrian{\raise 1.25pt
   \hbox to 3bp{
\hfill}\nobreak\ }
\def\solidsquare{\vrule height .9ex width .8ex depth -.1ex\nobreak\ }

\def\solidcclose{\drawline{10}{.5}\nobreak\raise
  0.5pt\hbox{$\bullet$}\drawline{10}{.5}\nobreak\ }

\def\solidsclose{\drawline{10}{.5}\nobreak\raise
  0.5pt\hbox{\solidsquare}\drawline{10}{.5}\nobreak\ }

\def\solidtclose{\drawline{10}{.5}\nobreak\raise
  0.5pt\hbox{\solidtrian}\drawline{10}{.5}\nobreak\ }

\def\solidcopen{\drawline{10}{.5}\nobreak\raise
  0.5pt\hbox{\circle}\drawline{10}{.5}\nobreak\ }

\def\solidsopen{\drawline{10}{.5}\nobreak\raise
  0.5pt\hbox{\square}\drawline{10}{.5}\nobreak\ }

\def\solidtopen{\drawline{10}{.5}\nobreak\raise
  0.5pt\hbox{\trian}\drawline{10}{.5}\nobreak\ }

\def\solidx{\drawline{10}{.5}\nobreak\raise
  0.5pt\hbox{\x}\drawline{10}{.5}\nobreak\ }

\begin{document}

\preprint{APS/123-QED}

\title{Heat transfer enhancement and reduction by  poylmer additives\\
 in turbulent Rayleigh Benard convection
}

\author{Yves Dubief}
 \email{yves.dubief@uvm.edu}
\affiliation{%
School of Engineering, Mechanical Engineering Program\\
 University of Vermont,
33 Colchester Ave, Burlington, VT 05405, USA
}%


\date{August 3, 2008}

\begin{abstract}
This letter confirms the existence of heat transfer enhancement (HTE) and reduction (HTR) in turbulent natural convection with polymer additives. HTE and HTR were numerically predicted by Benzi \textit{et al.}(PRL, \textbf{104} 024502, 2010) in homogenous turbulent convection, but experiments by Ahlers \& Nikolaenko(PRL, \textbf{104} 034503, 2010) in turbulent natural convection observed HTR only. Using direct numerical simulation of natural convection, the present study reconciles earlier numerical and experimental work on the basis of the dominant role of polymer length in the polymer dynamics in extensional flows.
\end{abstract}

\pacs{Valid PACS appear here}
\maketitle

In the turbulence research community, high-molecular weight polymer additives are well known for their ability to reduce turbulent drag\cite{sreenivasan2000odr,white2008map}. Their effect on turbulent heat transport is far less documented, even though polymer solutions may be of interest in the management of heat transfer via turbulence control. A recent experimental study\cite{ahlers2010effect} demonstrated that dilute solutions of polymers have the ability to produce heat transfer reduction (HTR). A direct numerical simulation, and shell model simulation of homogenous natural convection \cite{benzi2010effect} at $Pr=1$ reported a non-monotonic behavior of heat transfer as a function of the Weissenberg number $We$, ratio of the relaxation time scale of the polymer solution to the turbulent time scale of the flow. Using direct numerical simulation (DNS), heat transfer enhancement (HTE) was observed for all simulated $We\lesssim 1$ calculated from the flow time scale based on the rms of velocity fluctuations.  Benzi \textit{et al.} also used a shell model to expand the range of $We$ and predicted HTR at large $We$. Ahlers \& Nikolaenko\cite{ahlers2010effect} speculated that HTE might be the result of the absence of walls in the simulation. Hereafter the modification caused by polymer addition to heat transfer, measured by the Nusselt number $Nu$ the ratio of the convective to conductive heat fluxes,  is defined as $\text{HTE or HTR}=( Nu_\text{p}/Nu_\text{s}-1)\times100\;(\%)$  with respect to the Newtonian solvent heat transfer under the same temperature conditions

This letter reconciles Ahlers \& Nikolaenko\cite{ahlers2010effect}'s HTR measurements and Benzi \textit{et al.}\cite{benzi2010effect}'s predictions of HTE and HTR, by using direct numerical simulations of natural convection in a polymeric fluid between two infinite horizontal, isothermal walls. Additionally, the present study demonstrates  the importance of the polymer length $L$ in the selection of the regime of heat transfer (HTE or HTR), and identifies the specific polymer/flow interactions that lead to HTE or HTR.

Turbulent natural convection flows are simulated in a cartesian domain defined by the orthonormal vector base $(\vex{e}_x,\vex{e}_y,\vex{e}_z)$ where $x$, $y$ and $z$ are the two horizontal directions and the vertical direction, respectively. The components of the velocity vector $\vex{u}$ are $u$, $v$, and $w$ and are normalized by the  free-fall velocity, or convection velocity $U_c=\sqrt{\alpha gH\Delta}$, where $\alpha$, $g$, $H$ and $\Delta$  the fluid's coefficient expansion, gravity, distance between the two no-slip, isothermal, horizontal walls, and the temperature difference between the bottom (hot) and top (cold) walls. In all simulations, $H=1$ and $\Delta=1$. The Rayleigh number, ratio of buoyancy forces to thermal and momentum diffusive forces, $Ra=\alpha gH^3\Delta/(\nu\kappa)$ is low, $Ra=10^5$, yet within the turbulent regime.  The kinematic viscosity and thermal diffusivity of the solvent are identified as $\nu$ and $\kappa$, respectively. The Prandtl number is also fixed, $Pr=\nu/\kappa=7$ which is realistic for water. The Reynolds number based on $U_c$ and $H$ is therefore  $Re=Ra^{1/2}Pr^{-1/2}=119$. The flow is incompressible ($\vnabla\cdot\vex{u}=0$), periodic in horizontal directions and the buoyancy effect is simulated using the Boussinesq approximation, with the following transport equations for velocity, pressure $p$ and temperature $\theta$ and temperature fluctuations $\theta'$ around the hydrostatic temperature profile :
\begin{equation}
\pdert{\vex{u}}+(\vex{u}\cdot\vnabla)\vex{u}=-\vnabla p+\theta'\vex{e}_z+\frac{\beta}{Re}\nabla^2\vex{u}+\frac{1-\beta}{Re}\vnabla\cdot\tentau \label{eq:mom}
\end{equation}
\begin{equation}
\pdert{\theta}+(\vex{u}\cdot\vnabla)\theta=\frac{1}{PrRe}\nabla^2\theta\;.
\end{equation}

The parameter $\beta$ is the ratio of solvent viscosity to the zero-shear viscosity of the polymer solution and affects both the viscous stress and polymer stress terms in \eqnref{eq:mom}. The polymer stress tensor $\ten{T}$ is computed using the FENE-P (Finite Elastic Non-linear Extensibility-Peterlin) model:
\begin{equation}
\ten{T}=\frac{1}{We}\left(\frac{\ten{C}}{1-\text{tr}(\ten{C})/L^2}\label{eq:taup}
-\ten{I}\right)
\end{equation}
where the tensor $\ten{C}$ is the local conformation tensor of the polymer solution and $\ten{I}$ is the unit tensor. The properties of the polymer solution are $\beta$, the relaxation time, here based on the convection scales ($We=\lambda U_c/H$) and the maximum polymer extension $L$. The FENE-P model assumes that polymers may be represented by a pair of beads connected by a non-linear spring and defined by the end-to-end vector $\vex{q}$. The conformation tensor is the phase-average of the tensorial product of the end-to-end vector $\vex{q}$ with itself, $\ten{C}=\langle\vex{q}\otimes\vex{q}\rangle$ whose transport equation is
\begin{equation}
\pdert{\ten{C}}+(\vex{u}\cdot\vnabla)\ten{C}=\ten{C}(\vnabla\,\vex{u})+(\vnabla\,\vex{u})^\text{T}\ten{C}-\tentau \label{eq:C}
\end{equation}
On the lhs. of \eqnref{eq:C}, the first two terms are responsible for the stretching of polymers by hydrodynamic forces, while the third term models the internal energy that tends to bring stretch polymers to their least energetic state (coiled). The FENE-P model has demonstrated its ability to capture the physics of polymer drag reduction\cite{dubief2005nai,white2008map}. In the present work, we make the further assumption that the thermal conductivity is independent of the polymer concentration, which should be reasonable within the dilute approximation.

Eqs.~(\ref{eq:mom}-\ref{eq:C}) are solved using finite differences on a staggered grid, following \cite{dubief2005nai}. The code has been validated against existing databases of turbulent channel flows and natural convection simulations\cite{kerr2000prandtl}. All simulations are performed in a computational domain of dimensions $8H\times8H\times H$ and resolution $128\times128\times129$. The Newtonian, HTE$=+10\%$ ($L=25$. $We=10$), and HTR$=-30\%$ were repeated on domains with twice the spatial resolution and also doubled lateral dimensions with virtually no change in heat transfer.  

\begin{figure}
\includegraphics[width=0.35\textwidth]{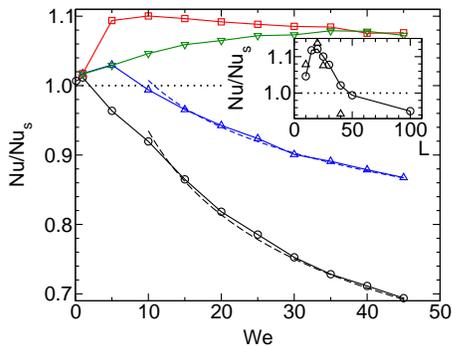}
\caption{\label{fig:Nu} Modification of the viscoelastic flow Nusselt number normalized by the solvent Nusselt number $Nu_s$. Main graph: $Nu/Nu_s$ as a function of the Weissenberg number $We$ for $L=10$ ($\bigtriangledown$), $L=25$ ($\Box$), $L=50$ ($\bigtriangleup$), and $L=100$ ($\bigcirc$). Power law fits of $Nu/Nu_s$(\dashed): $L=50$, $\propto We^{-0.1}$; $L=100$, $\propto We^{-0.2}$ . Insert: $Nu/Nu_s$ as a function of the polymer length $L$ for $We=10$ ($\bigcirc$) and $We=40$ ($\bigtriangleup$).
}
\end{figure}

In \figref{fig:Nu}, several simulations were used to map the heat transfer response of a wide range of relaxation time $0.1\leq We\leq45$, and polymer lengths $L=10,25,50,100$ with a few additional simulations for $L=15,20,30,40$ at $We=10$ and $L=25,40$ at $We=40$. Simulations at higher $We$ were found to require smaller time and space resolutions and will be discussed in subsequent publications. As predicted by the shell model of Benzi \textit{et al.}\cite{benzi2010effect}, HTE is observed at low We for nearly all polymer lengths (except for $L=100$), yet the present simulated enhancement (HTE$=+12\%$ for $L=25$ and $We=10$) is much more modest than the maximum observed in Benzi \textit{et al.}'s shell model of homogenous convection, $Nu/Nu_\text{s}\approx 6$, or their DNS where $Nu/Nu_\text{s}\approx 2.4$ with $L=30$. The absence of walls appears therefore to magnify the HTE ability of viscoelastic thermal convection flows. \figref{fig:Nu} also highlights the critical role of the polymer length on the heat transfer performance. Long polymers, $L\gtrsim 50$, show hardly any HTE. For the range of $We$ considered, the short polymer simulations show a decrease of the HTE effect, yet do not reach HTR for the range of considered $We$. The two HTR flows exhibit a power-law behavior at high $We$, with $Nu\propto We^{-0.1}$ and $We^{-0.2}$, for $L=50$ and 100, respectively. Although this result indicates a dependence of the power law exponent in $L$, the limited available data does not allow for a definitive conclusion. Combining the shell model and the scaling theory of Grossmann and Lohse\cite{grossmann2000scaling},  Benzi \textit{et al.}\cite{benzi2010effect} predicted the HTR behavior to $Nu\propto We^{-1}$. Since the shell model prediction were performed at much higher Rayleigh numbers, it may be assumed that Benzi's work may describe an asymptotic behavior of HTR.

\begin{figure}
\includegraphics[width=0.42\textwidth]{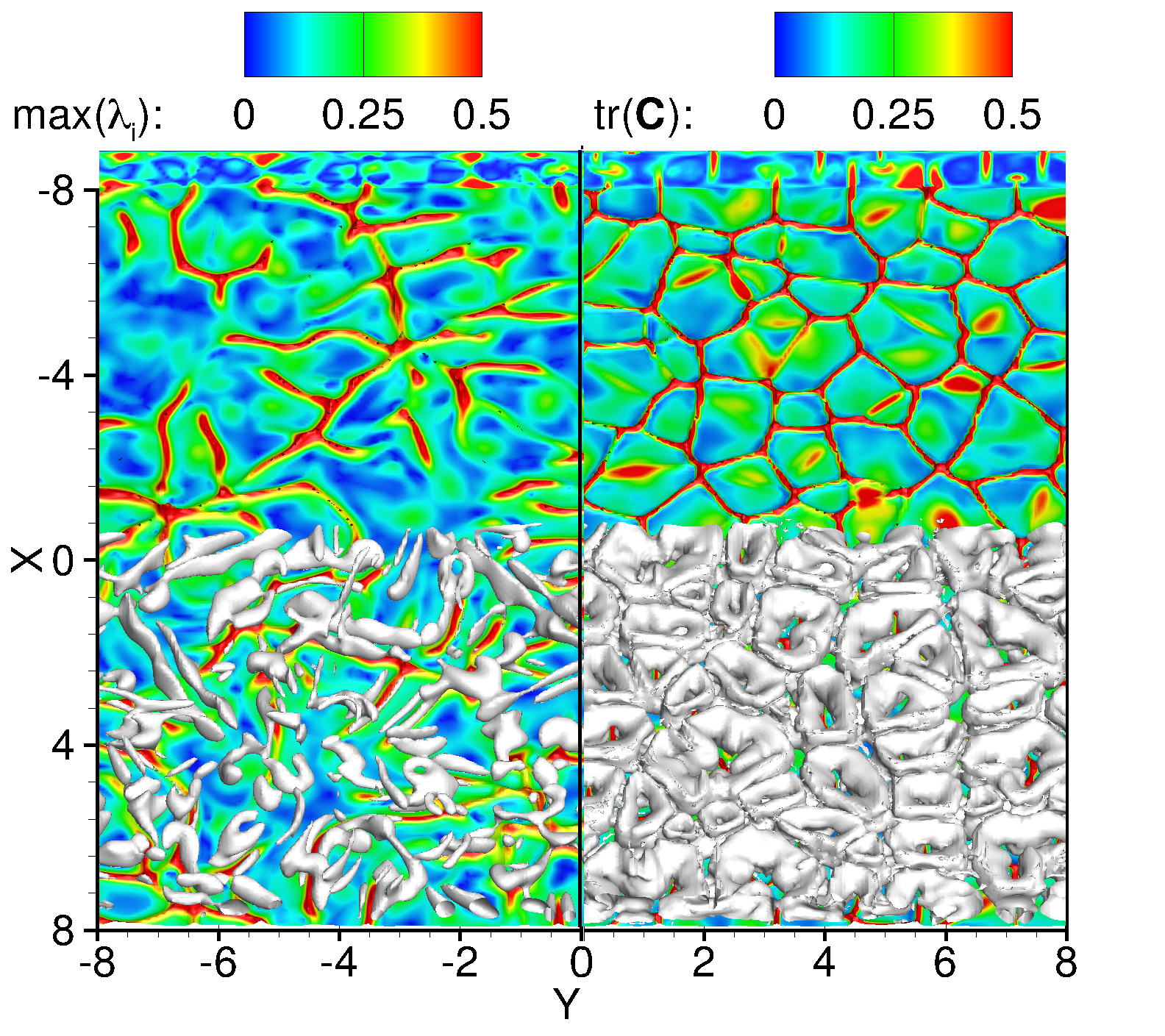}\\
\includegraphics[width=0.42\textwidth]{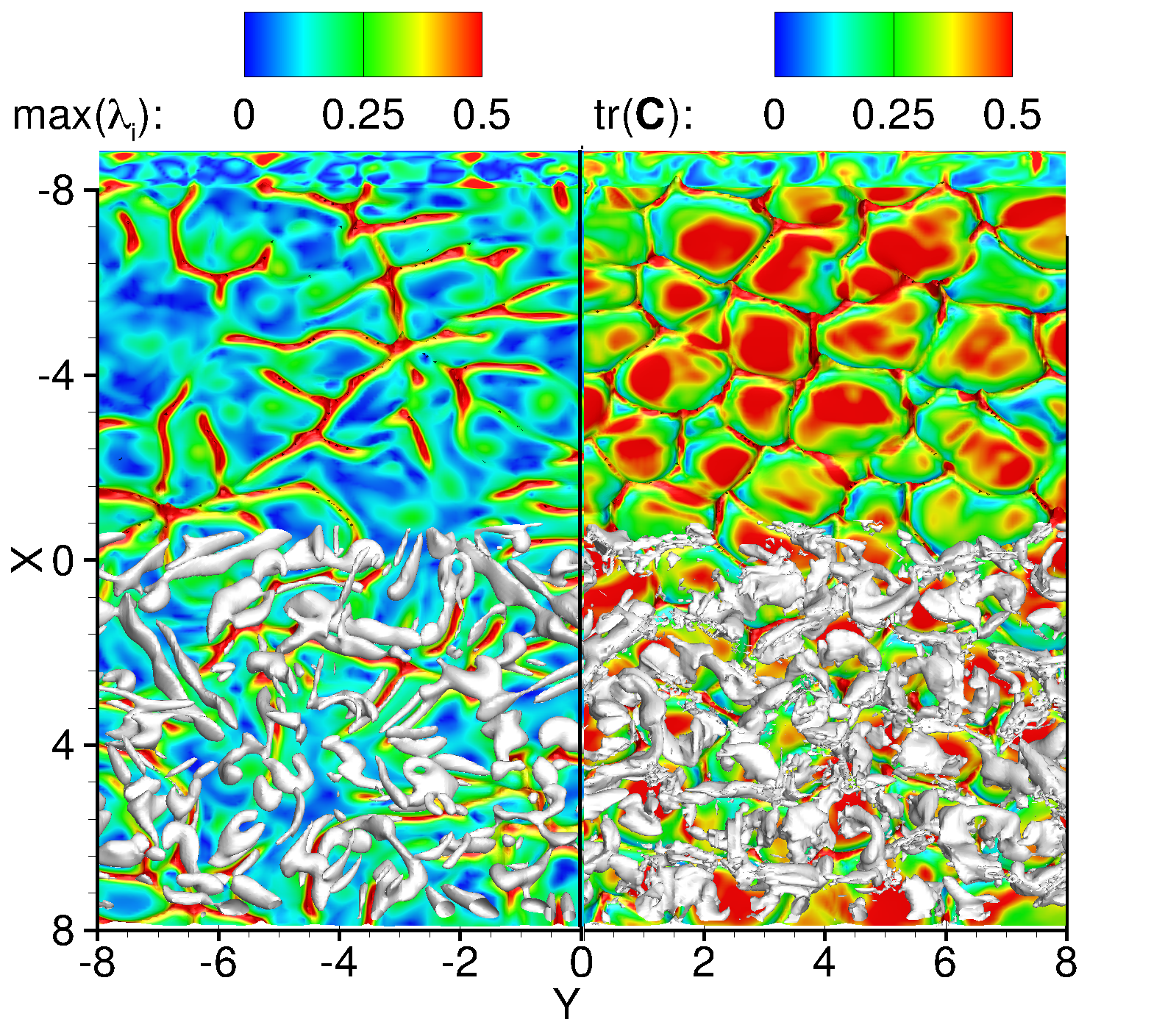}
\caption{\label{fig:viz} Juxtaposed snapshots of domain halves of the Newtonian flow ($-8\leq y/H\leq0$) and viscoelastic flows ($0\leq y/H\leq+8$) for HTE$=+11\%$ (top) and  HTR$=-30\%$ (bottom), looking down to the hot plate at a $45^\circ$ angle. The geometry of the convection cells is highlighted by an isosurface of temperature represents $\theta=0.85$ colored by the maximum magnitude of eigenvalues of the local velocity gradient tensor $\lambda^*$ (Newtonian flow) or the normalized trace of the conformation tensor $\text{tr}(\ten{C})/L^2$ (viscoelastic flows). The panels at $x/H=-8$ are vertical contour slices of the flow domain using $\lambda^*$ and $\text{tr}(\ten{C})/L^2$. White isosurfaces of the second invariant of the velocity gradient tensor $Q$ show the topology regions dominated by rotation\cite{dubief2000cvi} over $0\leq x/H\leq8$: Newtonian, $Q=0.1$; viscoelastic, $Q=0.01$.}
\end{figure} 

The topological differences between Newtonian, HTE and HTR flows are depicted in \figref{fig:viz}. The shape of convection cells is identified by the ridges observed in an instantaneous isotherm at $\theta=0.85$. For the Newtonian flow, these ridges are highlighted by contours of $\lambda^*=\max\vert\lambda^r_i\vert$, $\lambda^r_i$ is the real part of the eigenvalues of the velocity gradient tensor $\vnabla\vex{u}$. Terrapon \textit{et al.}\cite{terrapon2004sps} introduced  $\lambda^*$ as a measure of the ability of turbulent  flows to stretch polymer molecules. It should be noted that regions of large $\lambda^*$ are also regions or uni- or biaxial-extensional flows\cite{chong1990gct}. For viscoelastic flows, the isotherm is colored by the local polymer stretch, calculated as $\text{tr}(\ten{C})/L^2$. The topology of convection cells is dramatically modified by the presence of the polymers, with the emergence of highly organized convection cells. The typical horizontal length scale of convection cells is measured using  the first negative minimum of the  radial correlation function of vertical velocity fluctuations in planes parallel to the walls (not shown). This length scale  drops from 3.2 in the Newtonian flow to 0.87 in the HTR flow and  0.7 in the HTE flow. The structure of regions of rotation-dominated flows is shown by positive isosurfaces of the second invariant $Q$ of the velocity gradient tensor\cite{dubief2000cvi}. The $Q$-criterion is sensitive to the intensity of vortices and is adjusted to identify vortical structures in the core of the Newtonian flows with $Q=0.1$, arising from shear layer instability in plumes. Such small scales vortices are absent in the HTE flow, and, with a much lower threshold $Q=0.01$, the structure of convection cells is isolated. The vertical slice of polymer stretch contours shown in \figref{fig:viz} indicates sustained stretching over most of the plumes' vertical extent. The absence of core turbulence is not surprising since polymers have been shown to stabilize shear layers\cite{azaiez1994lsf}. The HTR flow does not exhibit as much coherence as HTE. In fact convection cells identified by $Q=0.01$ isosurfaces appear to break down in smaller structures. The vertical slices show a much rapid extinction of polyerm stretch away from the origin of the plume, as well as more horizontal motion of polymer stretch. Using two-dimensional simulations, the present author has isolated in plumes a polymer-driven instability at large $We$ and $L$, which is the likely cause of the observed small scales. Since the initial study suggests that this instability is not a major component of the HTR mechansim, this matter will be discussed in a future publication.  Lastly, \figref{fig:viz} shows much lower polymer stretch in the boundary layers of HTE convection cells than HTR.

\begin{figure}
\includegraphics[width=0.35\textwidth]{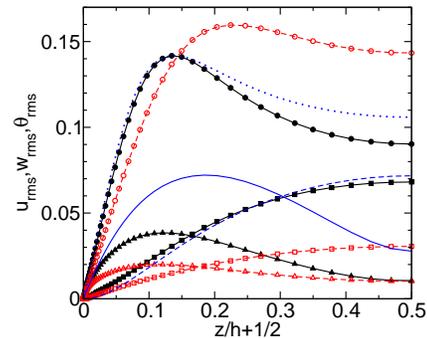}
\caption{\label{fig:P} Profiles of rms of velocity and temperature fluctuations in the lower half of the computational domain. Newtonian, HTE and HTR flows are denoted by lines without symbols, closed and open symbols, respectively. $u_\text{rms}$: \solid, $\bigtriangleup$; $w_\text{rms}$: \dashed, $\Box$; $\theta_\text{rms}$: \dotted, $\bigcirc$.
}
\end{figure}

\figref{fig:P} plots the vertical distributions of the rms of temperature, horizontal and streamwise velocity fluctuations. The respective thickness of the momentum $\delta_u$ and thermal $\delta_\theta$ boundary layers is estimated by the altitude of the maxima of $u_\text{rms}$ and $\theta_\text{rms}$\cite{kerr2000prandtl}. Newtonian and HTE temperature and vertical velocity fluctuations are very close, especially in the boundary layer region. The main difference is observed in the horizontal velocity, with a reduction of $\delta_u$ ($\delta^\text{HTE}_u\approx0.12<\delta_\theta\approx0.14<\delta^\text{Newt}_u\approx0.18$) and a reduction of velocity magnitude. The HTR flow significantly departs from the Newtonian flow for all quantities shown in \figref{fig:P}. The momentum thickness is virtually identical to that of HTE but the thermal boundary layer is much thicker ($\delta_\theta\approx0.22$). Temperature fluctuations are increased in plumes but velocity fluctuations are significantly reduced in both plumes and boundary layers. The absence of core turbulence in HTR and HTE flows leads to convection cell-driven flows which is reflected in the fact that $w\text{rms}\approx 2\times u_\text{rms}$, or the  the intensity of plume velocity fluctuations are the sum of velocity fluctuations in the boundary layers of two adjacent convection cells. 

\begin{figure}
\includegraphics[width=0.35\textwidth]{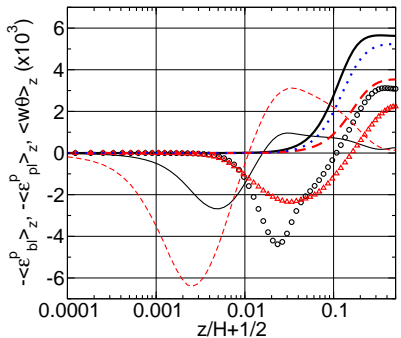}
\caption{\label{fig:E} Vertical profiles of the elastic energy contribution to the Nusselt number as described in \eqnref{eq:Nu}. $-\langle\varepsilon^p_{bl}\rangle_z$: \solid, HTE; \dashed, HTE. $-\langle\varepsilon^p_{pl}\rangle_z$: $\bigcirc$, HTE; $\bigtriangleup$, HTR. Vertical heat flux $\langle w\theta\rangle_z$: \boldsolid, HTE; \dotted, Newtonian; \bolddashed, HTR.
}
\end{figure}
\figref{fig:E} relates the heat flux $\langle w\theta\rangle_z=\kappa\Delta/H(Nu-1)$ to the elastic energy $\langle\varepsilon^p\rangle_z=\langle \vex{u}\cdot((1-\beta)/Re\vnabla\cdot\ten{T})\rangle_z$, where $\langle\cdot\rangle_z$ denotes the averaging over time and homogenous direction at altitude $z$.  Following Grossmann \& Lohse\cite{grossmann2000scaling}, the focus of the discussion is the relation:
\begin{equation}
\frac{\nu^3}{H}(Nu-1)RaPr^{-2}=\langle\varepsilon^u_{bl}\rangle_V+\langle\varepsilon^u_{pl}\rangle_V-\langle\varepsilon^p_{bl}\rangle_V-\langle\varepsilon^p_{pl}\rangle_V
\label{eq:Nu}
\end{equation}
where $\varepsilon^u$ is the TKE dissipation rate, $\langle\cdot\rangle_V$ the averaging over time and the entire flow volume, and the subscripts $bl$ and $pl$ denotes the respective contribution of boundary layers and plumes. For clarity, \figref{fig:E} plots only $-\langle\varepsilon^p_{bl}\rangle_z$ and $-\langle\varepsilon^p_{pl}\rangle_z$, calculated from the horizontal and vertical contributions of $\varepsilon^p$. In the HTE flow, the volume average of the  plume elastic energy, $-\langle\varepsilon^p_{pl}\rangle_V=0.86$ dominates that of boundary layers $-\langle\varepsilon^p_{bl}\rangle_V=0.17$. The HTR flow shows the inverse behavior with $-\langle\varepsilon^p_{pl}\rangle_V=0.32$ and $-\langle\varepsilon^p_{bl}\rangle_V=0.46$. The explanation for the different distribution of elastic energy is first found in plumes. The Newtonian flow, via contours of $\lambda^*$, shows that the base of plumes hosts the largest extensional velocity gradients, and polymeric fluids indeed exhibit the most stretch in these regions (\figref{fig:viz}). As polymers travel in the plume, the velocity gradient drops rapidly, similar to the relaxation of a stretched polymer arising from the sudden cancellation of an extensional flow studied by Doyle \textit{et al.}\cite{doyle1998relaxation}. These authors demonstrated that the first normal stress decreases at a rate of $L\sqrt{We/t}$. Replacing time by travel distance in the plume, this decay rate can be observes in the vertical slice of the contours of  $\text{tr}(\ten{C})/L^2$ of  \figref{fig:viz}, showing much more rapid relaxation in HTR than HTE. Short polymers are therefore able to sustain larger contributions over longer plume lengths. The second difference  is the significant stretch in boundary layers at HTR, while almost negligible at HTE. The negative (positive) elastic contribution of $\langle\varepsilon^p_{bl}\rangle_z$ ($-\langle\varepsilon^p_{bl}\rangle_z$) for $z/H+1/2>0.01$ is the result of a large increase of extensional viscosity driven by accelerating fluid in this region. Below the extensional region, boundary layers are shear dominated and thus experience the effect of shear thinning as shown by positive (negative) elastic contribution of $\langle\varepsilon^p_{bl}\rangle_z$ ($-\langle\varepsilon^p_{bl}\rangle_z$). Extensional viscosity for a FENE-P fluid increases dramatically with increasing $L$ and $We$\cite{bird1987dynamics}, which leads to the reduction of horizontal velocity fluctuations in the HTR flow (\figref{fig:P}). 

In conclusion, the commonality in the present HTE and HTR flows is the absence of core turbulence, which implies that both flows are driven by convection cells alone. The present HTE simulations and the homogenous convection simulations of Benzi \textit{et al.}\cite{benzi2010effect} demonstrate that the interactions of polymers with plumes is inherently of HTE nature for short polymer length. Long polymers, via extensional viscosity in boundary layers and rapid relaxation in plumes, cause HTR as observed in Ahlers \& Nikolaenko\cite{ahlers2010effect}'s experiment with high molecular weight polymers at much higher Rayleigh numbers. The extrapolation of the proposed models to high Reynolds number should be valid on the basis that polymers have been shown to reduce turbulence intensity and mixing in free shear\cite{vaithianathan2007polymer} and isotropic\cite{perlekar2006manifestations} turbulence, which should be the dominant flow patterns in core turbulence of natural convection  at high $Ra$.

\begin{acknowledgments}
The computational resources provided by the Vermont Advanced Computing Center which is supported by NASA ( NNX 06AC88G) are gratefully acknowledged.
\end{acknowledgments}

\bibliography{Natural_convection_polymers}

\end{document}